# МОДУЛЬ В ЧЕРТЕЖЕ КАК ОСНОВА ТЕХНОЛОГИИ РАЗРАБОТКИ ПРОБЛЕМНО-ОРИЕНТИРОВАННЫХ РАСШИРЕНИЙ САПР

Мигунов В.В., г.Казань

Для САПР, направленных на создание чертежей, разработка специализированных проблемно-ориентированных расширений оправдана в случаях, когда за счет их применения трудоемкость ввода данных существенно ниже трудоемкости непосредственного черчения с помощью графического ядра САПР, включая проведение необходимых расчетов. Например, когда требуются трудоемкие расчеты или когда нормативные требования к чертежу порождают большое количество графических изображений по малому количеству исходных данных. В обоих случаях наиболее предпочтительна параметрическая генерация чертежа по исходным данным. Для решения задач такого класса разработана технология, основанная на совместном хранении в одном элементе чертежа, называемом "Модуль", как исходных данных, так и результатов параметрической генерации.

Модуль включает видимую в чертеже совокупность геометрических элементов и невидимое в чертеже параметрическое представление моделируемого объекта. Третья часть проблемно-ориентированного расширения - процедуры работы с модулем, они помещаются в программный код САПР. Ниже приводятся примеры моделируемых объектов, отвечающие потребностям САПР реконструкции химического предприятия, где чертежи выпускаются по СПДС:

условные графические обозначения приборов и исполнительных механизмов в схемах автоматизации. По текстовым признакам генерируется стандартное обозначение, зависящее от них. В модулях хранится специфицирующая информация об изделиях, выбранная в электронных каталогах, для генерации спецификаций;

монтажный чертеж трубопровода, генерируемый по осевой ломаной и установкам (диаметр, способ стыковки в углах и др.);

чертежи строительной подосновы в плане и в разрезе обеспечивают автоматический тираж строительных осей и их обозначений, размеров и др.;

аксонометрические схемы трубопроводных систем. В модуль включается даже библиотека используемых в нем условных графических обозначений. Обеспечивается автоматическая привязка обозначений к трубам таким образом, чтобы выносные линии размеров шли вдоль координатных осей, перенос веток трубопровода и др.;

табличный модуль. Универсальная модель табличных конструкторских документов, позволяющая корректно размещать данные в произвольные таблицы из электронных каталогов с автоматической фильтрацией, пересчетом единиц измерения и др.;

профили наружных сетей водоснабжения и канализации. Автоматически создается не только сам профиль, но и таблица основных данных;

проект молниезащиты зданий и сооружений. Позволяет наглядно работать с сечениями зон защиты, рассчитывать варианты геометрии которых вручную очень трудоемко. Автоматически создается таблица основных данных.

Параметрические представления объектов в модулях вместе со специализированным кодом расширения САПР позволяют реализовывать самые разные модели объектов и методов их разработки. Специализация модулей легко опознается в чертеже, и возникает возможность многоэтапной разработки объектов проектирования и использования объектов - прототипов. Высокая структурированность параметров обеспечивает быстрый доступ к ним для различной обработки. Например, в модуль аксонометрической схемы можно автоматически вставить оси строительной подосновы, лишь выбрав модуль подосновы в чертеже. Легко автоматизируется сбор сведений при генерации спецификаций, при контроле дублирования позиционных обозначений..

Комплект параметров модулей ряда объектов может записываться на диск (без геометрических элементов), порождая информационную среду проектирования в виде библиотек прототипов. При выборе комплекта для чтения с диска геометрия генерируется в режиме on-line, и проектировщик легко ориентируется в прототипах.

Как элемент чертежа модуль подчиняется обычным правилам: его можно удалить, подвинуть, растянуть и т.д., к его геометрическим элементам возможны привязки, он может быть помещен в графическую библиотеку. Внутри модуля для входящих в него геометрических элементов хранятся габариты по зонам чертежа, позволяющие игнорировать при выводе на экран ненужные элементы. Все элементы модуля лежат на одном слое, но могут иметь разные типы линий и цвет.

При задании параметрического представления используется как стандартный пользовательский интерфейс (меню, формы ввода...), так и специализированный для черчения и корректировки. Технология реализована в САПР реконструкции химического предприятия TechnoCAD GlassX Подробнее: technocad.narod.ru.